\documentstyle[iopconf2]{article}

\newcommand{\ra}{\mbox{$\rightarrow$}}
\newcommand{\pipm}{\mbox{$\pi^+\pi^-$}}
\newcommand{\optbar}[1]{\shortstack{{\tiny (\rule[.4ex]{1em}{.1mm})} 
  \\ [-.7ex] $#1$}}
\newcommand{\abb}{\mbox{$A(B_q\ra \overline{B_q})$}}
\newcommand{\abf}{\mbox{$A(\overline{B_q}\ra f)$}}
\newcommand{\Gqf}{\mbox{$\Gamma_{qf}(\tau)$}}
\newcommand{\agqf}{\mbox{$\overline{\Gamma_{qf}}(\tau)$}}
\newcommand{\Eq}[1]{Eq.~(\ref{eq#1})}

\newcommand{\beq}{\begin{eqn}}
\newcommand{\eeq}{\end{eqn}}
\newcommand{\nl}{\label{eq\theequation}  \\ \addtocounter{eqns}{1}}
\newcounter{eqns}
\renewcommand{\theequation}{\thesection.\theeqns}
\newenvironment{eqn}{\addtocounter{eqns}{1} \begin{equation}
  \label{eq\theequation}}{\end{equation}}
\renewcommand{\theequation}{\thesection.\theeqns}

\begin{document}

\input BoxedEPS.tex
\SetOzTeXEPSFSpecial
\HideDisplacementBoxes 

\begin{flushright} NSF-PT-97-4 \end{flushright}
\title{Future probes of the origin of CP violation\footnote[2]{To appear in the 
Proceedings of the Workshop "Beyond the Desert---Accelerator and Non-Accelerator 
Approaches", held in Tegernsee, Germany, June 1997.}}

\author{Boris Kayser}

\affil{National Science Foundation \\ 4201 Wilson Boulevard,
  Arlington, VA  22230  USA}

\beginabstract
We review what has been learned about CP violation in the K system. It is natural
to hypothesize that the observed CP-violating effects are caused by the Standard
Model weak interaction. We describe the stringent future test of this hypothesis
via experiments on the $B$ system. Then, we see how new physics beyond the
Standard Model could be revealed by this test.
\endabstract

\section{What do we already know? \label{sect1}}
\setcounter{eqns}{0}
Before discussing the future probes of the origin of CP violation, 
let us briefly recall some of the things we already know.

All laboratory CP-violating effects seen so far occur in the neutral 
$K$ system. We know that the neutral $K$ mass eigenstates, $K_{\mbox {Short}}
(K_{S})$ and $K_{\mbox {Long}} (K_{L})$, are not CP eigenstates, as they would
be if the  world were CP invariant. Rather, they are CP admixtures given by
\addtocounter{eqns}{1}
\begin{eqnarray} 
  |K_S\rangle &=& |K_1 \rangle + \epsilon |K_2 \rangle \nonumber \\
  |K_L\rangle &=& |K_2 \rangle + \epsilon |K_1 \rangle . \label{eq\theequation}
\end{eqnarray}
Here, $K_1$ and $K_2$ are, respectively, the CP-even and CP-odd 
eigenstates of CP, and $|\epsilon| = \Or(10^{-3})$. We know also that
both $K_L$ and $K_S$ can decay to the CP-even final state $\pi^+\pi^-$. 
The ratio of their decay amplitudes, $\eta_{+-} \equiv \langle\pi^+\pi^-|T|
K_L\rangle / \langle\pi^+\pi^-|T|K_S\rangle$, has the measured value \cite{r1}
\beq
\eta_{+-} = [(2.285 \pm 0.019) \times\ 10^{-3}] \,\e^{i(43.56\pm0.56)^0} . 
\eeq

The CP-violating decay $K_L \ra \pi^+\pi^-$ can, in principle,
result from two effects. First, it can occur because, as described by Eqs. (1.1), 
$K_L$ has a small $K_1$ component, which can decay to \pipm\ without 
violating CP. This type of CP violation, which results from CP-noninvariance 
of the $K^{0}-\overline{K^0}$ mixing amplitudes which make $K_S$
and  $K_L$ what they are, is referred to as ``indirect CP violation.'' 
In addition, it may be that even the $K_2$ component of $K_L$ can decay to 
$\pi^+\pi^-$. This type of CP violation, in which a decay amplitude itself
violates CP, is called ``direct CP violation''.\ \cite{r2}
 
The CP-violating decay of $K_L$ to the CP-even final state $\pi^0\pi^0$ has
also been seen, and it has been found that $\eta_{00} \equiv
\langle\pi^0\pi^0|T|K_L\rangle / \langle\pi^0
\pi^0|T|K_S\rangle$ has the value [3] 
\beq
  \eta_{00} = [(2.275 \pm 0.019) \times\ 10^{-3}] \,\e^{i(43.5 \pm
  1.0)^\circ} .  \eeq

If direct CP violation may be neglected compared with the indirect kind
in $K \ra\  \pi\pi$ (that is, if we may take
$\langle\pi\pi|T|K_2\rangle  \cong\ 0$), then it follows from Eqs.~(1.1) that 
\beq  \eta_{+-} = \eta_{00} = \epsilon  .  \eeq
Comparing Eqs.~(1.2) and (1.3), we see that, indeed,$\eta_{+-}$ and 
$\eta_{00}$ are at least very close to being equal. In addition, 
assuming CPT, one predicts that \cite{r4}
\beq  \arg (\epsilon) = (43.46 \pm 0.08)^\circ  .  \eeq
This predicted phase of $\epsilon$ is in excellent agreement with the 
measured phases of $\eta_{+-}$ and $\eta_{00}$. Thus, we have
several pieces of evidence that \Eq{1.4} holds; that is, several
indicators that direct CP violation $\ll$ indirect CP violation in
$K\ra\pi\pi$.

A further fact that we know is that the CP-violating asymmetry
\beq
\delta_{\pi\ell\nu} \equiv \frac{\Gamma(K_L \rightarrow\ \pi^-\ell^+\nu)
- \Gamma(K_L \rightarrow\ \pi^+\ell^-\overline{\nu})}{''\hspace{.5in} 
+ \hspace{.5in}''} \eeq
 has the value [3]
\beq   \delta_{\pi\ell\nu} = (3.27 \pm 0.12) \times\ 10^{-3}  . \eeq
Now, assuming the validity of the $\Delta S = \Delta Q$ rule, $K_L%
\ra \pi^-\ell^+\nu$ comes only from the $K^0$ component of the
$K_L$, and $K_L\ra \pi^+\ell^-\overline{\nu}$ comes only from
the $\overline{K^0}$ component. If we assume also that there is no
significant direct CP violation in $K \ra\ \pi \ell \nu$, then 
\beq \left| \langle \pi^-\ell^+\nu |T| K^0\rangle \right| = 
   \left| \langle \pi^+\ell^-\overline{\nu} |T| \overline{K^0}\rangle
   \right|    .  \eeq
It is then easily shown from Eq. (1.1) for $K_L$ and from the familiar
expressions for $K_1$ and $K_2$ in terms of $K^0$ and
$\overline{K^0}$ that, for small $\epsilon$,
\beq  \delta_{\pi\ell\nu} = 2\, \Re (\epsilon)  .  \eeq
Now, if direct CP violation is indeed negligible in $K\ra \pi
\pi$, so that $\epsilon = \eta_{+-}$ [cf. Eq. (1.4)], then Eq. (1.9)
implies that
\beq  \delta_{\pi\ell\nu} = 2 \,\Re (\eta_{+-})  .  \eeq
From the experimental value (1.2) for $\eta_{+-}$, we find that
\beq  2 \,\Re (\eta_{+-}) = (3.31 \pm 0.04) \times 10^{-3}  ,  \eeq
in excellent agreement with the measured value (1.7) of
$\delta_{\pi\ell\nu}$. Thus, \Eq{1.10} does hold within errors,
providing evidence that direct CP violation is indeed very small in 
$K \ra\ \pi \ell \nu$, and adding to the evidence that it is also
very small in $K\ra \pi\pi$.

\section{What is the origin of CP violation? \label{sect2}}
\setcounter{eqns}{0}
All laboratory CP-violating effects observed to date have been seen in kaon
decays. Kaon decays are due to the weak interaction. Thus, the most obvious
candidate for the source of CP violation is the weak interaction itself.

The weak interaction, as described by the Standard Model (SM), can produce CP
violation only through complex phases in the Cabibbo-Kobayashi-Maskawa (CKM)
quark mixing matrix
\beq V = \left( \begin{array}{lll}
		V_{ud} & V_{us} & V_{ub} \\
		V_{cd} & V_{cs} & V_{cb} \\
		V_{td} & V_{ts} & V_{tb} 
	\end{array} \right)  . \eeq

Complex phases in V could not  have any physical consequences (such as CP
violation) if there were only two quark generations.\ \cite{r5}\, Furthermore,
in $K$ decay and mixing, the leading SM processes (e.g., the process $s \ra u
+ \overline{u} + d$ for $K$ decay) do not involve the $t$ or $b$ quarks. Hence,
speaking approximately, these processes ``do not know'' that the third quark
generation exists. Thus, again speaking approximately, $K$ decay and mixing
cannot violate CP through these leading SM processes, but only through SM
processes with smaller amplitudes. As a result, if complex phases in the SM quark
mixing matrix V are the source of CP violation, we expect this violation to be
small in $K$ decays, as observed.\ \cite{r6}

In the SM, we also expect that direct CP violation $\ll$
indirect CP violation, both in neutral $K \ra \pi\ell\nu$ and in neutral $K \ra
\pi\pi$, as observed. However, the relative smallness of direct CP violation in
these decays does not point uniquely to the SM. For example, if CP violation
arises, not from the SM weak interaction, but from a ``superweak'' interaction
that affects $K^0-\overline{K^0}$ mixing but makes only negligible contributions
to $K$ decay amplitudes, then once again we expect that direct CP violation
$\ll$ indirect CP violation in $K \ra \pi\ell\nu$ and $K\ra \pi\pi$.

While the SM expectation for $K \ra \pi\pi$ is that  Eq. (1.4) should hold
approximately, it would take an accident for $(|\eta_{+-}|^2 - |\eta_{00}|^2) /
(|\eta_{+-}|^2 + |\eta_{00}|^2)$ to be much smaller than $10^{-4}$.\ \cite{r7}
Thus, vigorous efforts are being made at Fermilab, CERN, and Frascati to detect
and measure a nonvanishing, if small, difference between $\eta_{+-}$ and
$\eta_{00}$.

The SM picture of CP violation is compatible, not only with all existing
information from the kaon system, but also with the bounds on the electric dipole
moments of various elementary particles. To be sure, it appears that CP violation
coming from CKM phases which produce their effects through the physics of the SM
cannot account for the baryon asymmetry of the universe.
Thus, this asymmetry may be pointing to physics beyond the SM. However, it is
thought that this asymmetry developed when the universe was still at a
temperature at or above $M_W$. For CP violation at energies well below $M_W$,
CKM phases acting within the SM remain a very plausible explanation. The
hypothesis that these phases, acting in this way, are indeed the origin of
low-energy CP violation will be tested cleanly and in detail during the next
ten to fifteen years. The test will be carried out mostly through experiments
on the $B$ system, but will also entail some important experiments on the $K$
system. Physics beyond the SM could be revealed through failure of the SM of CP
violation to pass the test posed by all these experiments.
  
\section{Testing the SM of CP violation in B decays \label{sect3}}
\setcounter{eqns}{0}
\subsection{What is there to measure? \label{ssect3.1}}
In $B$ decays, some of the anticipated CP-violating effects can cleanly probe
the phases of various products of CKM elements, thereby testing whether these
complex phases are indeed behind CP violation. There are only four independent
phases of CKM products, which may be taken \cite{r8} to be the quantities 
\addtocounter{eqns}{1}
\begin{eqnarray}
	\alpha &\equiv& \arg\ (-\frac{V_{td}V_{tb}^*}{V_{ud}V_{ub}^*})  ,
	  \label{eq\theequation}  \\ \addtocounter{eqns}{1}
	\beta  &\equiv& \arg\ (-\frac{V_{cd}V_{cb}^*}{V_{td}V_{tb}^*})  ,
	  \label{eq\theequation} \\ \addtocounter{eqns}{1}
	\chi   &\equiv& \arg\ (-\frac{V_{cb}V_{cs}^*}{V_{tb}V_{ts}^*})  ,
	  \label{eq\theequation} \\ \addtocounter{eqns}{1}
	\chi'  &\equiv& \arg\ (-\frac{V_{us}V_{ud}^*}{V_{cs}V_{cd}^*})  .
	  \label{eq\theequation}  
\end{eqnarray}
The phases $\alpha$ and $\beta$ may be pictured as two of the angles in the 
``$db$ unitarity triangle'', which expresses pictorially the SM unitarity
constraint that the $d$ and $b$ columns of the CKM matrix must be orthogonal:
\beq  V_{ud}V_{ub}^* + V_{cd}V_{cb}^* + V_{td}V_{tb}^* = 0  .   \eeq
This triangle is shown in Fig.~1. In a similar way, $\chi$ is an angle in the
$sb$ unitarity triangle, which expresses the orthogonality of the $s$ and $b$ CKM
columns, and $\chi'$ is an angle in the $ds$ triangle, which expresses the
orthogonality of the $d$ and $s$ columns. The angles $\alpha$ and $\beta$ may
both be quite large, but, given what we already know about the CKM elements,
$\chi$ is at most a few percent of a radian, and $\chi'$ at most a few
milliradians.
\begin{figure}		
	\hSlide{1.5in}
	\tBoxedEPSF{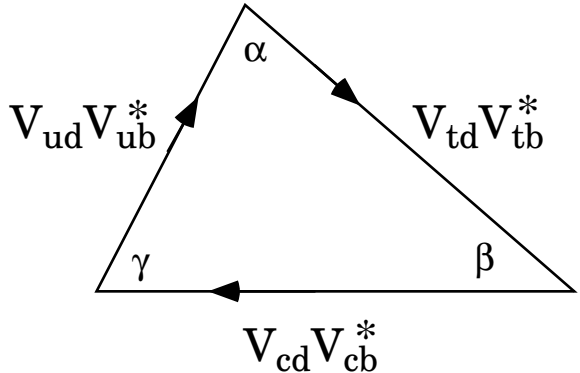  scaled 1000}
	\caption{The $db$ unitarity triangle, drawn somewhat schematically,
	expressing the orthogonality of the $d$ and $b$ columns of the CKM matrix.
	The phases $\alpha$ and $\beta$ defined by Eqs.~(\ref{eq3.1}) and
	(\ref{eq3.2}) are the two angles so labelled in this triangle. The third
	angle in the triangle is called $\gamma$. \label{f1}}
\end{figure}

The CP asymmetry in a given B decay mode will probe the phase of some
phase-convention-independent product of CKM elements. One can show [8] that if
$\varphi$ is the phase of such a product, then, $\bmod\,\pi$,
\beq  \varphi\ = n_{\alpha}\alpha +  n_{\beta}\beta + n_{\chi}\chi + 
 n_{\chi'}\chi'  ,  \eeq
where $n_{\alpha}$, $n_{\beta}$, $n_{\chi}$, and $n_{\chi'}$ are integers. For
the $B$ decay mode studied in a typical CP experiment, each of these integers
is $0$, $\pm 1$, or $\pm 2$, so that the relation between $\varphi$ and the
underlying angles $\alpha$, $\beta$, $\chi$, and $\chi'$ is almost trivial.
Thus, the experiments that will study CP violation in the B system may be
thought of as, among other things, measurements of the angles $\alpha$, 
$\beta$, $\chi$, and $\chi'$. Interestingly, once these four angles are
determined, the entire CKM matrix, including the magnitude and phase of each of its elements, 
follows from them.\ [8] Thus, in principle, CP
violation in the $B$ system probes the entire content of the CKM matrix. However,
it is the independent phases $\alpha$, $\beta$, $\chi$, and $\chi'$ of CKM
products which are the probed quantities which are simply related to what will be
observed.

As we shall see shortly, when the CP asymmetry in some $B$ decay mode probes a
CKM phase $\varphi$ which is one of the small angles $\chi$ or $\chi'$, the CP
asymmetry itself is small and, consequently, hard to measure. Indeed, when the
phase probed is $\chi'$, the CP asymmetry is only $\Or(10^{-3})$, and
experiments to study such a small asymmetry may never be practical in the $B$
system. However, experiments to measure the (hopefully) large angles $\alpha$,
$\beta$, and $\gamma \equiv \pi-\alpha-\beta$ in the $db$ triangle are
being very actively developed, and experiments to determine the $\Or(10^{-2}$
radians) angle $\chi$ are being contemplated as well.

The program to test the SM of CP violation via experiments on B decays may be
summarized as follows:
\begin{enumerate}
	\item Measure the four independent phases $\alpha$, $\beta$, $\chi$, and 
	$\chi'$ of CKM products. If the smallest of these, $\chi'$, is beyond
	reach, at least measure $\alpha$, $\beta$, and $\chi$. Focus first on
	$\alpha$ and $\beta$, since these phases may both be large.
	\item To see whether the SM provides a consistent picture of CP-violating
	phenomena or leads to inconsistencies which point to physics beyond the SM,
	overconstrain the system as much as possible. To do so---
	\begin{enumerate}
		\item Measure, if possible, CP asymmetries in different decay modes
		which, {\em if} the SM of CP violation is correct, all yield the same
		phase angle ($\beta$, for example). See whether these asymmetries
		actually yield the same numerical result. 		
		\item Measure {\em independently} the angles $\alpha$, $\beta$, and 
		$\gamma$ in the $db$ triangle, and see whether these angles actually
		add up to $\pi$.
		\item Measure the lengths of the sides of the $db$ triangle (via
		experiments on non-CP-violating effects such as decay rates and neutral
		B mixing). See whether the interior angles implied by the measured 
		lengths agree with those inferred directly from CP-violating
		asymmetries.
	\end{enumerate}
\end{enumerate}

\subsection{How will the phases of CKM products be cleanly measured?%
\label{ssect3.2}}
The techniques through which CKM phase information can be extracted from B
decays have been extensively discussed in the literature.\ [9] Here, we shall
only recall some highlights.

With some notable exceptions, the $B$ decays that can yield clean CKM phase
information are of the neutral B mesons, $B_d\ (=\overline{b}d)$ and 
$B_s\ (=\overline{b}s)$. Each of these mesons mixes significantly with its
antiparticle. In the SM, the $B_q$-$\overline{B_q}\ (q = d$ or $s$) mixing is
due largely to the box diagram in Fig.~2. This diagram obviously imparts to the
mixing amplitude \abb\ the CKM phase [10]
\beq  \arg_{CKM} \abb\ = 2 \, \arg\,(V_{tq}V_{tb}^*)  .  \eeq
\begin{figure}			
	\hSlide{1.25in}
	\tBoxedEPSF{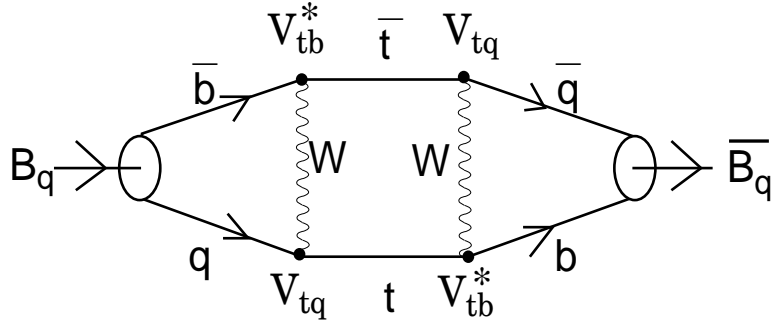  scaled 1000}
	\caption{The SM box diagram for $B_q$-$\overline{B_q}$ mixing. \label{f2}}
\end{figure}

Owing to mixing, there are two ways in which a neutral B that is initially a pure
$B_q$ can decay to some final state $f$: The $B_q$ can decay directly to $f$. Or,
it can convert via $B_q$-$\overline{B_q}$ mixing into a $\overline{B_q}$, and
then the $\overline{B_q}$ decays to $f$. The amplitude for the direct decay,
$A(B_q\ra f)$, interferes with that for the mixing-induced decay, \abb\,\abf.
Suppose that in each of the amplitudes $A(B_q\ra  f)$, \abb, and \abf, some one
Feynman diagram, which of course is proportional to some product of CKM elements,
dominates. Then the interference between the direct-decay and mixing-induced
paths from the initial $B_q$ to the final state $f$ probes the single CKM phase
$\varphi_{qf}$ which is the relative CKM phase of the two interfering paths. That
is,
\beq
	\varphi_{qf} = \arg_{CKM}\,\left[ \frac{A(B_q\ra f)}{\abb\,\abf} \right]  .
\eeq

A particularly simple situation arises when $f$ is a CP eigenstate. Suppose a 
$B$ meson is at proper time $\tau = 0$ a pure $B_q$ (pure $\overline{B_q}$). Let
the rate for this $B$ to decay to a CP eigenstate $f$ at proper time $\tau$ be
denoted by \Gqf\ [\agqf]. One finds that, if each of the amplitudes appearing
in Eq. (3.8) is dominated by one Feynman diagram, then \Gqf and \agqf are given
by [9]
\beq
	\optbar{\Gamma}_{qf} (\tau) \propto\ \e^{-\Gamma_q\tau} \left[ 1 
	\stackrel{\raisebox{-.8ex}{+}}{\raisebox{-1.1ex}{{\tiny (}$-${\tiny )}}}
	\eta_f\sin\varphi_{qf}\sin(\Delta 	M_q\tau) \right] .
\eeq
Here, $\Gamma_q$ is the width of the two mass eigenstates of the $B_q-
\overline{B_q}$ system (we neglect the expected $\sim 20\%$ width difference of
the two $B_s$ mass eigenstates). The quantity $\Delta M_q$ is the positive
mass difference between the two $B_q$ mass eigenstates, and $\eta_f$ is the CP
parity of $f$.

Since \Gqf and \agqf are the rates for two CP-mirror-image processes, the
asymmetry between these rates, 
\beq
	{\cal A}_{qf}(\tau) \equiv\ \frac{\Gqf - \agqf}{\Gqf + \agqf} =
	\eta_f \sin\varphi_{qf} \sin (\Delta M_q\tau)  ,
\eeq
is a violation of CP. Note that, assuming $\eta_f$ and $\Delta M_q$ to be
known, a measurement of ${\cal A}_{qf}(\tau)$ would yield a clean
determination of $\sin\varphi_{qf}$.

When each of the amplitudes in \Eq{3.8} is dominated by one Feynman diagram,
proportional to some product of CKM elements, then $\varphi_{qf}$ is obviously
the phase of some product of CKM elements. Thus, the CP asymmetry (\ref{eq3.10})
cleanly probes the phase of a CKM product. Note from Eqs. (3.8)--(3.10) that,
within the SM, it is only the CKM phase, and not the magnitude, of the mixing
amplitude \abb\ to which CP-violating asymmetries are sensitive. More generally,
if \abb\ should contain a contribution from beyond the SM, this contribution
would affect CP-violating asymmetries only if it modified the {\em phase} of
\abb.\ [11]

As \Eq{3.10} makes clear, when $\varphi_{qf}$ is one of the large angles
$\alpha$, $\beta$, or $\gamma$, the CP asymmetry ${\cal A}_{qf}$ will be large,
quite unlike the asymmetry $\delta_{\pi\ell\nu}$ in $K_L$ decay. However, as 
\Eq{3.10} also makes clear, when $\varphi_{qf}$ is one of the small angles $\chi$ or
$\chi'$, ${\cal A}_{qf}$ is correspondingly small.

How does one identify the $B$ decay modes for which $\varphi_{qf}$ is a
particular CKM phase angle of interest, such as $\alpha$ or $\beta$? This
question is most easily answered if we make Wolfenstein's approximation [12] to
the CKM matrix. In this approximation, in a common phase convention, the only
CKM elements which depart significantly from being real are $V_{ub}$ and
$V_{td}$. Then, from Eqs. (3.1)--(3.4) and the constraint that 
$\gamma =\pi-\alpha-\beta$,
\addtocounter{eqns}{1}
\begin{eqnarray}
	\alpha  &\cong& \pi + \arg\,(V_{td}) + \arg\,(V_{ub})  \:, \nl
	\beta   &\cong& -\arg\,(V_{td})  \:, \nl
	\gamma  &\cong& -\arg\,(V_{ub})  \:, \nl
	\chi    &\cong& 0  \:,  \nl
	\chi'   &\cong& 0  \:.  \label{eq\theequation}
\end{eqnarray}
Thus, from \Eq{3.7},
\beq  \arg_{CKM}\, A(B_d \rightarrow \overline{B_d}) \cong\ -2\beta  \:, \eeq 
while
\beq  \arg_{CKM} A(B_s \rightarrow \overline{B_s}) \cong 0  \:. \eeq
From these relations and \Eq{3.8} we see that if, for example, we would like 
$\varphi_{qf}$ to be $\sim\beta$, we may choose a $B_d$ decay mode (where
mixing involves $\beta$) in which the CKM elements appearing in the decay
amplitudes $A(B_q \ra f)$ and \abf are $\sim$ real (so that there are no
further CKM phases). An example, $B_d \ra\ D^+D^-$, where $\varphi_{qf} =
2\beta$, is shown in Table 1. Similarly, if we wish $\varphi_{qf}$ to be
$\sim\alpha$, we may choose a $B_d$ decay mode in which the decay amplitudes
involve $V_{ub}$. An example, $B_d \ra\ \pi^+\pi^-$, where $\varphi_{qf} =
-2\alpha$, is given in Table 1. If we wish $\varphi_{qf}$ to be
$\sim\gamma$, we may select a $B_s$ decay mode (where mixing introduces no
phase) in which the decay amplitudes involve $V_{ub}$. An example \cite{r13},
$B_s\ra  D^{+}_sK^-$, where $\varphi_{qf} = \gamma$, is shown in Table 1. In this
example, the final state is not a CP eigenstate, and the decay rate is not
described by \Eq{3.9}. However, the CKM phase probed is still the relative CKM
phase of the direct-decay and the mixing-induced paths to the final state, as
described by \Eq{3.8}.

To consider the small angle $\chi$, we must go beyond Wolfenstein's
approximation. Table \ref{t1} lists a decay mode, $B_s\ra \psi\phi$, which
probes $\chi$. Indeed, from Eqs.~(3.3), (3.8), and (3.7), and the final column of
Table \ref{t1}, we find that in $B_s\ra \psi\phi$, $\varphi_{qf} = -2\chi$.
\begin{table}		
\centering 
	\caption{Illustrations of decay modes that probe $\beta$, $\alpha$,
	$\gamma$, and $\chi$. In the second column is shown the diagram which
	dominates $A(B_q \ra f)$, and in the third the one which
	dominates \abf. A wavy line in any of these diagrams denotes a $W$ boson.
	In the final column is given the CKM factor to which $A(B_q\ra f) / \abf$
	is proportional.  \label{t1}} 
\begin{tabular}{lcc}  \topline
Decay mode			& $A(B_q\ra f)$ \hspace*{.75in} \abf\ 			&    
  \shortstack[l]{Decay CKM \\ factor}  \\  \midline
\begin{minipage}[t]{.8in}
$B_d \ra\ D^+ D^-$	 \vspace*{1.05in}  \\  
$B_d \ra\ \pipm$	 \vspace*{1.05in}  \\  
$B_s \ra\ D^+_s K^-$ \vspace*{1.05in}  \\
$B_s \ra\ \psi \phi$ \vspace*{1.05in}	
\end{minipage}	& 	  
\raisebox{.3in}{\tBoxedEPSF{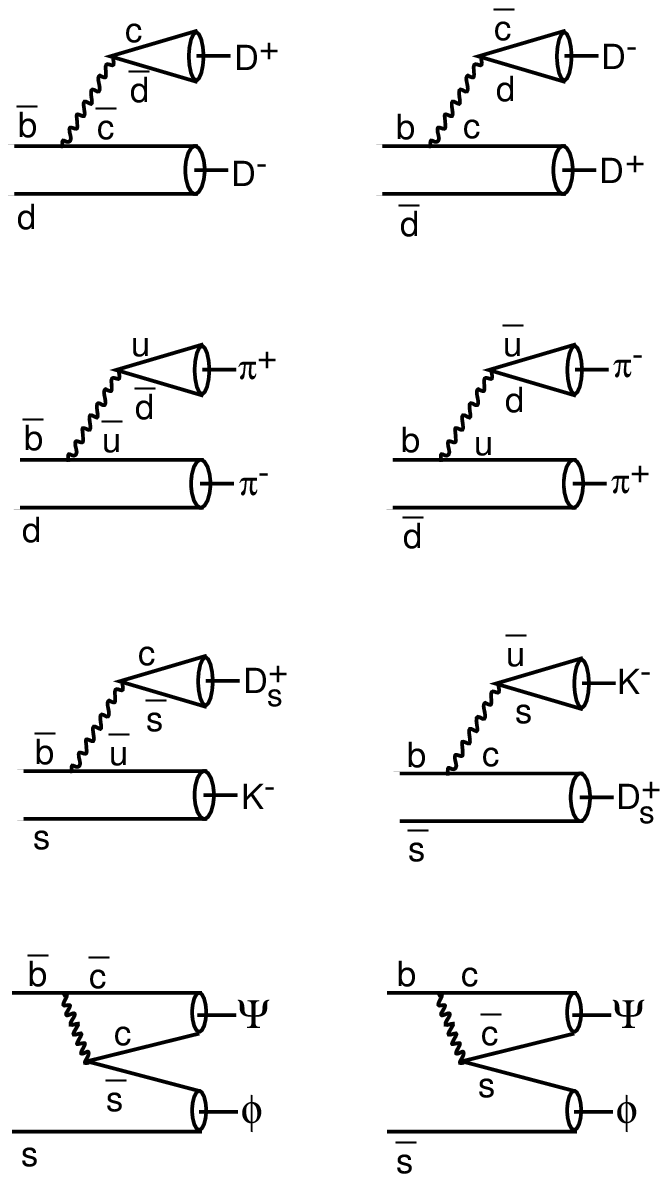  scaled 950}}	&
\begin{minipage}[t]{.6in}
   $\displaystyle \frac{V^*_{cb}V_{cd}}{V_{cb}V^*_{cd}}$  \vspace*{.9in} \\
   $\displaystyle \frac{V^*_{ub}V_{ud}}{V_{ub}V^*_{ud}}$  \vspace*{.9in} \\
   $\displaystyle \frac{V^*_{ub}V_{cs}}{V_{cb}V^*_{us}}$  \\ [.9in]
   $\displaystyle \frac{V^*_{cb}V_{cs}}{V_{cb}V^*_{cs}}$   
\end{minipage}  \\ \bottomline
\end{tabular}
\end{table}
\subsection{New Wrinkles \label{ssect3.3}}
We would like to mention several interesting recent developments concerning the
test of the SM of CP violation in $B$ decays.

\begin{figure}[tb] 		
	\hSlide{1.1in}
    \tBoxedEPSF{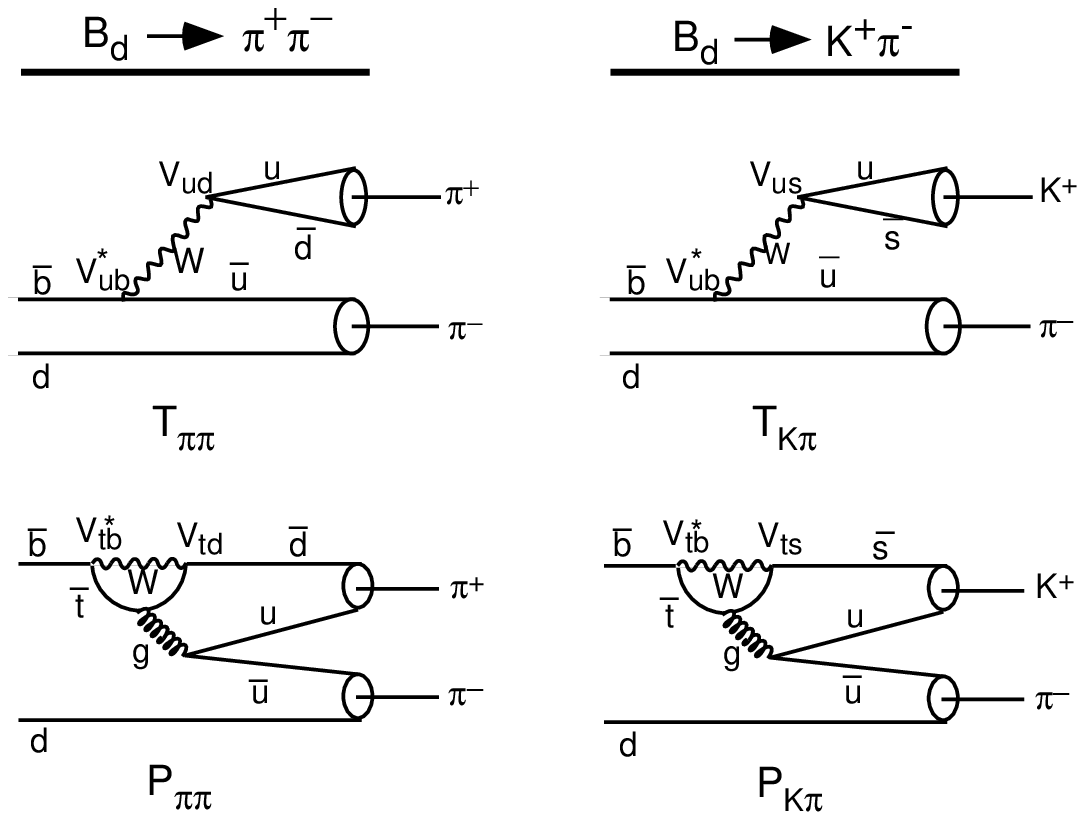  scaled 1000}
	\caption{Decay diagrams for $B_d\ra \pipm$ and $B_d\ra K^+ \pi^-$. 
	\label{f3}}
\end{figure}

First, data from the CLEO collaboration at Cornell suggest that in
$\optbar{B_d}\ra \pipm$, the decay mode where information on the angle $\alpha$
will almost certainly first be sought, the assumption that one diagram dominates
$A(B_d\ra\pipm)$, while other contributions to this decay amplitude may be safely
neglected, is invalid. The relevant data are the branching ratio \cite{r14}
\beq
	BR(B_d \rightarrow  K^+ \pi^-) = (1.5 \raisebox{-.5ex}{ \scriptsize 
	\shortstack[r]{+0.52  \\  \m0.42}}) \times 10^{-5} , \eeq  
and the $(90\% CL)$ bound \cite{r14}
\beq
	BR(B_d \rightarrow  \pipm) < 1.5  \times 10^{-5}  .  \eeq
Now, the decay amplitudes for $B_d\ra K^+ \pi^-$ and $B_d\ra \pipm$ are expected to
receive contributions from the diagrams in Fig.~\ref{f3}.
For each decay mode, there is a tree diagram, labelled T in Fig.~\ref{f3}, and
a ``penguin'' diagram, labelled P. In $B_d\ra \pipm$, the tree diagram is expected
to dominate. The major difference between the tree diagram for $B_d\ra \pipm$,
$T_{\pi\pi}$, and the one for $B_d\ra K^+ \pi^-$, $T_{K\pi}$, is that $V_{ud}$
in the former is replaced by $V_{us}$ in the latter. Thus,
\beq  \left| T_{K\pi} \right|  \simeq \lambda \left| T_{\pi\pi} \right|  ,  \eeq
with $\lambda \equiv V_{us} / V_{ud} = 0.22$. Similarly, the major difference
between the penguin diagram for $B_d\ra \pipm$, $P_{\pi\pi}$, and the one for
$B_d \ra K^+ \pi^-$, $P_{K\pi}$, is that $V_{td}$ in the former is replaced by
$V_{ts}$ in the latter, Thus, estimating that $|V_{td}/V_{ts}|\sim\lambda$, we
have
\beq    \left| P_{K\pi} \right|  \simeq 
			\frac{1}{\lambda} \left| P_{\pi\pi} \right|  .  \eeq

The experimental results \Eq{3.18} and (\ref{eq3.19}) suggest that $BR(B_d\ra
\pipm)$ $\leq BR(B_d\ra K^+\pi^-)$. In view of relations (\ref{eq3.20}) and
(\ref{eq3.21}), this in turn suggests that $P_{\pi\pi}$ cannot be entirely
negligible compared to $T_{\pi\pi}$. Indeed, if we use Eqs.~(\ref{eq3.20}) and
(\ref{eq3.21}), but allow for various possible values of the relative phase of
$P_{\pi\pi}$ and $T_{\pi\pi}$, and various possible values of that of
$P_{K\pi}$ and $T_{K\pi}$, we find that
\beq
	\left| \frac{P_{\pi\pi}}{T_{\pi\pi}} \right| \approx 
	\left\{ \begin{array}{ll}
		0.1 - 0.4 &\mbox{if } BR(B_d\ra \pipm) = \0\, BR(B_d\ra K^+\pi^-) \\
		0.2 - 0.6 &\mbox{if } BR(B_d\ra \pipm) = 
			\frac{1}{2} BR(B_d\ra K^+\pi^-) \\
		0.3 - 0.9 &\mbox{if } BR(B_d\ra \pipm) = 
		\frac{1}{4} BR(B_d\ra K^+\pi^-)
	\end{array} \right.	
\eeq
Clearly, one cannot safely assume that the penguin diagram may be neglected
relative to the tree diagram in $B_d\ra \pipm$.

Fortunately, a technique was developed long ago to deal with a possibly
non-negligible penguin contribution to $B_d\ra \pipm$.\ \cite{r15} This
technique exploits the fact that the penguin and tree diagrams for $B\ra\pi\pi$ have different
isospin properties, and requires the experimental study of several
isospin-related $B\ra\pi\pi$ decays.

A second recent development we would like to mention is the suggestion that
experimental study of ``cascade mixing'' could yield the sign of $\cos 2\beta$,
and thereby eliminate the anticipated discrete ambiguities in the angles
$\alpha$, $\beta$, and $\gamma$. \cite{r16}

To test the SM of CP violation, one would like to determine the angles in the
unitarity triangles, especially those in the $db$ triangle of Fig.~\ref{f1}.
However, as \Eq{3.10} illustrates, CP asymmetries determine only trigonometric
functions of these angles, leaving the angles themselves discretely ambiguous.
As the time when the B-system CP experiments will be done approaches, means for
eliminating these discrete ambiguities are being developed.

Most likely, the first CKM phase quantities to be determined will be $\sin
2\beta$ (from $\optbar{B_d}\ra\psi K_S)$, $\sin 2\alpha$ (from $\optbar{B}\ra
\pi\pi$), and $\cos 2\gamma$ (perhaps from $B^\pm\ra DK^\pm$, as discussed
shortly). If we assume that $\alpha$, $\beta$, and $\gamma$ are the three angles
in a triangle, then a knowledge of $\sin 2\alpha$, $\sin 2\beta$, and $\cos
2\gamma$ will fix these three angles completely, except for a two-fold ambiguity.
That ambiguity would be resolved if one could determine either Sign $(\cos
2\alpha)$ or Sign $(\cos 2\beta)$.\ \cite{r17}

\begin{figure}			
	\hSlide{1.15in}
	\tBoxedEPSF{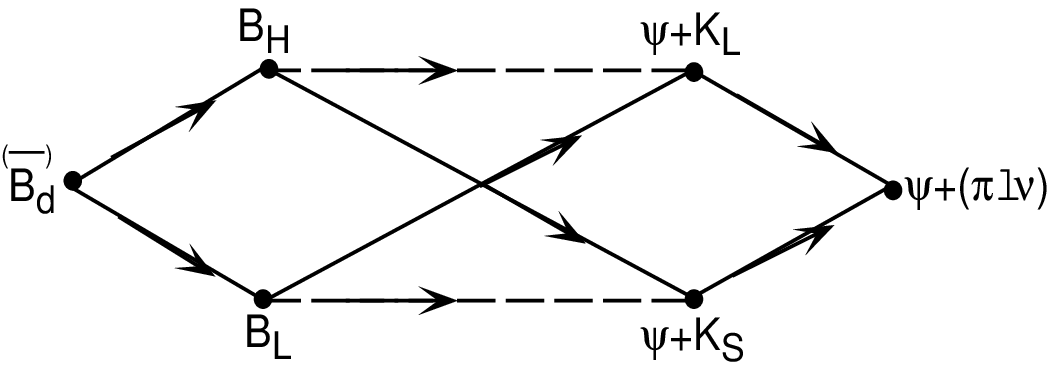  scaled 1000}
	\caption{The paths through which the decay chain $\protect\optbar{B_d} \ra
	\psi 	+ 	K 	\ra \psi + (\pi\ell\nu)$ can proceed. \label{f4}}
\end{figure}

The sign of $\cos 2\alpha$ might be found through analysis of the three-body
decays $\optbar{B_d}\ra\pipm\pi^0$.\ \cite{r18} The sign of $\cos 2\beta$
could be found \cite{r16} by studying the decay chain
\beq	\optbar{B_d} \rightarrow  \psi + K \rightarrow  \psi + (\pi\ell\nu) .
\eeq 
In this chain, neutral $B$ mixing before the primary decay is followed by neutral
$K$ mixing after it. We refer to this as ``cascade mixing''.\ \cite{r19} This
decay chain is sensitive to both $\sin 2\beta$ and $\cos 2\beta$, even though
$\optbar{B_d} \ra  \psi + K_S$ depends only on $\sin 2\beta$. To see why, let us
consider Fig.~\ref{f4}, which shows the paths through which the decay chain
(\ref{eq3.23}) can occur.
In Fig.~\ref{f4}, the mass eigenstates of the $B_d-\overline{B_d}$ system are
labelled $B_H$ ($B_{\mbox{Heavy}}$) and $B_L$ ($B_{\mbox{Light}}$). As
Fig.~\ref{f4} indicates, a $B$ which starts out as a pure $B_d$ or a pure 
$\overline{B_d}$ has both a $B_H$ and a $B_L$ component. Either of these
components can decay to either $\psi+K_L$ or $\psi +K_S$. Subsequently,
either the $K_L$ or $K_S$ can decay to $\pi\ell\nu$, and the amplitudes for
these decays are very comparable. Thus, there are four
paths from the initial \optbar{B_d} to the final state, $\psi+(\pi\ell\nu)$.
The amplitudes for all of these paths depend on the same CKM phase angle,
$\beta$, but they depend on $\beta$ in different ways. In the limit where CP
is conserved (so that $\beta$ and all other CKM phases vanish), all the
intermediate states in Fig.~\ref{f4}, $B_H$, $B_L$, $\psi K_L$, and $\psi
K_S$, are CP eigenstates. In particular, $CP(B_H) = CP(\psi K_S) = -1$, while 
$CP(B_L) = CP(\psi K_L) = +1$. Thus, the decays $B_H\ra \psi K_L$ and 
$B_L\ra \psi K_S$, represented by dashed lines in Fig.~\ref{f4}, connect
states which in the CP-conserving limit are of opposite CP parity. Therefore, the
amplitudes for these decays must vanish in this limit, and one finds that, in
particular, they are proportional to $\sin \beta$. By contrast, the decays 
$B_H\ra \psi K_S$ and $B_L\ra \psi K_L$, represented by solid lines in 
Fig.~\ref{f4}, connect states which in the CP-conserving limit are of the same
CP parity. Thus, the amplitudes for these decays are expected to survive in this
limit, and one finds that they are proportional to $\cos \beta$. Now, from
Fig.~\ref{f4}, we see that the decays \optbar{B_d}$\ra \psi K_S$ involve only two
paths, one through $B_H\ra \psi K_S$ and one through $B_L\ra \psi K_S$. It is the
interference between these two paths that leads to the CP asymmetry in
\optbar{B_d} $\ra \psi K_S$. Since $A(B_H\ra \psi K_S) \propto \cos \beta$,
while $A(B_L\ra \psi K_S) \propto \sin \beta$, the interference between them
is proportional to $\cos \beta \sin \beta$, or to $\sin 2\beta$. This is why 
\optbar{B_d} $\ra \psi K_S$ probes only $\sin 2\beta$. In contrast, in the decay
chain (\ref{eq3.23}), there are the four paths shown in Fig.~\ref{f4}, and all
of them interfere. Since $A(B_H\ra \psi K_S)$ and $A(B_L\ra \psi K_L)$ are
both $\propto \cos \beta$, the interference between them is $\propto \cos^2 \beta$.
Similarly, the interference between $A(B_H\ra \psi K_L)$ and $A(B_L\ra \psi
K_S)$ is $\propto \sin^2 \beta$. Obviously, a suitable linear combination of
$\cos^2 \beta$ and $\sin^2 \beta$ will yield $\cos 2\beta$. This is why
$\optbar{B_d} \ra  \psi + K \ra  \psi + (\pi\ell\nu)$ can determine $\cos
2\beta$.\ \cite{r20}

The event rate for $\optbar{B_d} \ra  \psi + K \ra  \psi + (\pi\ell\nu)$ is
such that, hopefully, this mode could be used to extract Sign $(\cos 2\beta)$
at a hadron facility, although the extraction may not be feasible at an $e^+
e^-$ $B$ factory. 

The final development we would like to mention is a refinement \cite{r21} of the
technique \cite{r22} for extracting $\gamma$ from the {\em charged} $B$ decay
chain
\beq  B^+ \rightarrow  K^+ + \optbar{D} \rightarrow K^+ + (f_{CP})  \eeq
and its CP conjugate. In the chain (\ref{eq3.24}), $f_{CP}$, a CP eigenstate,
can come either from an intermediate $D$ or a $\overline{D}$. It is easily
found that, in the Wolfenstein approximation and phase convention leading to
Eqs.~(\ref{eq3.11})--(\ref{eq3.15}), $A(B^+\ra K^+ D)\propto\e^{i\gamma}$,
while $A(B^+\ra K^+ \overline{D})$ and the \optbar{D} decay amplitudes have no
appreciable CKM phases. Thus, the interference between the amplitude for 
$B^+ \ra  K^{+}+D \ra K^{+}+(f_{CP})$ and that for $B^+ \ra  K^{+}+ 
\overline{D} \ra K^{+}+(f_{CP})$ probes $\gamma$. However, to extract clean
information on $\gamma$ from measured rates for (\ref{eq3.24}) and its CP
conjugate, one needs to know $|A(B^+\ra K^+ D)|$ and
$|A(B^+\ra K^+ \overline{D})|$, which clearly affect the rates. These magnitudes
can in principle be determined from the branching ratios for $B^+\ra K^+ D$ and 
$B^+\ra K^+ \overline{D}$. However, to measure these branching ratios, one must
be able to identify a $D$ and a $\overline{D}$ in the final state. It has
recently been noticed that, in the case of the $D$, this may prove very
difficult.\ \cite{r21} For example, the D from $B^+\ra K^+ D$ cannot be
identified via its decay to $K^-\pi^+$. For, $\overline{D}$ decay to
$K^-\pi^+$, while highly suppressed, occurs as well, and the fact that  
BR$(B^+ \ra  K^{+}\overline{D}) \gg $ BR$(B^+ \ra  K^+ D)$ has the consequence
that BR$(B^+ \ra K^+ \overline{D})\, $BR$(\overline{D} \ra  K^- \pi^+)$ is
expected to be comparable to
BR$(B^+ \ra  K^+ D)\, $BR$(D \ra  K^- \pi^+)$. Identification of a
$\overline{D}$ has no such problem, so it may be that one will be able to
determine $|A(B^+\ra K^+ \overline{D})|$, but not $|A(B^+\ra K^+ D)|$.

With this possibility in mind, it has been suggested \cite{r21} that, instead
of studying the decay chain (\ref{eq3.24}), one study {\em two} chains:
\beq  B^+ \ra  K^+  + \optbar{D} \ra K^+ + (f_1)  ,  \eeq
\beq  B^+ \ra  K^+  + \optbar{D} \ra K^+ + (f_2)  ,  \eeq
and their CP conjugates. Here, $f_1$ and $f_2$ are final states into which the
intermediate $D$ or $\overline{D}$ has decayed. These final states $f_i,
i=1,2$, are not CP eigenstates, but are chosen so that the interfering
amplitudes for $B^+ \ra  K^++D \ra K^++(f_i)$ and $B^+ \ra  K^++\overline{D}
\ra K^++(f_i)$ are comparable, maximizing any CP-violating effect.

In the SM, each of $B^+ \ra  K^+ \overline{D}$ and $B^+ \ra  K^+ D$ is
dominated by a single diagram, so that
\beq  |A(B^+ \ra K^+\overline{D})| = |A(B^- \ra K^- D)| \equiv a  ,   \eeq
and
\beq  |A(B^+ \ra K^+D)| = |A(B^- \ra K^- \overline{D})| \equiv b  .   \eeq
We assume that $a$ will be determined from BR$(B^+ \ra  K^+ \overline{D})$, while
the smaller $b$ will be unknown. With the branching ratios for $D$ decay
assumed known, the branching ratios $\rho_j, j=1,\ldots ,4$, for the decay
chains (\ref{eq3.25}), (\ref{eq3.26}), and their CP conjugates will then depend
on four unknowns: $b$, the strong-interaction phase of the amplitude for the chain
(\ref{eq3.25}), that of the amplitude for the chain
(\ref{eq3.26}), and the CKM phase angle $\gamma$. Thus, a measurement of the
four $\rho_j$ will yield information on $\gamma$.

\section{How can physics beyond the SM affect CP violation in the B system? 
\label{sect4}} 
\setcounter{eqns}{0}
By serving as a detailed, stringent test of the SM of CP violation, the study
of CP-violating asymmetries in $B$ decays can perhaps reveal evidence of
Physics Beyond the Standard Model (PBSM).

The feature of the $B$ system which is most susceptible to being affected by
PBSM is probably $B-\overline{B}$ mixing. This is due to the fact that the
mixing amplitude is very small, as we see, for example, from the tininess of
the mass splitting to which it leads:
\beq  \left. \frac{\Delta M}{M} \right|_{B_d} = 0.6 \times 10^{-13}  .  \eeq
Due to the smallness of the mixing amplitude, small effects from PBSM have a
chance to be visible.

As already mentioned in Sec.~\ref{ssect3.2}, a contribution to the mixing
amplitude \abb\ from PBSM can affect CP violation only if
it changes the {\em phase} of \abb. Now, there are combinations of measurements
that could reveal that the weak (i.e., non-strong-interaction) phase of \abb\
does not have its SM value, given by \Eq{3.7}. However, other combinations of
measurements would devilishly hide this fact. Let us see why this is so.

Suppose that, while the $B$ decay amplitudes are unaffected by PBSM, the mixing
amplitude \abb\ contains a contribution from  PBSM which changes its weak phase
to the SM value plus an offset $2\theta_q$:
\beq  \arg \abb  =  2 \arg\, (V_{tq}V_{tb}^*) + 2 \theta_q\:;\mbox{  q = d,s.} 
\eeq 
The weak phases probed by several popular $B$ decay modes are then modified as
described in Table \ref{t2}. Now, one important test of the SM of CP violation
will be to see whether the angles $\alpha, \beta, \gamma$ extracted from $B$
decays satisfy the constraint
\beq    \alpha  + \beta  + \gamma  =  \pi  ,  \eeq
as they must if they are actually the CKM phase angles in the $db$ unitarity
triangle of Fig.~\ref{f1}.\ \cite{r23} Suppose the values of $\alpha$, $\beta$,
$\gamma$, and $\chi$ inferred from $B$ experiments are $\tilde{\alpha}$,
$\tilde{\beta}$, $\tilde{\gamma}$, and $\tilde{\chi}$, respectively. If 
$\alpha$, $\beta$, and $\gamma$ are extracted from the first three processes
listed in Table \ref{t2}, then as this Table shows, $\tilde{\alpha} = \alpha +
\theta_d$, $\tilde{\beta} = \beta - \theta_d$, and $\tilde{\gamma} = \gamma$.
Thus, while the measured angles $\tilde{\alpha}$ and $\tilde{\beta}$ are not
the true CKM phase angles $\alpha$ and $\beta$, the measured angles
nevertheless satisfy
\beq  \tilde{\alpha} + \tilde{\beta} + \tilde{\gamma} = \pi  ,  \eeq
thereby concealing the presence of PBSM.\ \cite{r24}

\begin{table}		
	\caption{The phase quantities actually measured by popular decay modes in 
	the presence of PBSM. \label{t2}}  
	\centering 
	\[ \begin{array}{lcc} \topline
	\mbox{Process}	& \shortstack{Measures \\ in the SM} &
	  \shortstack{Actually measures when \\ PBSM is present}  \\  \midline
	\optbar{B_d}\ra\pipm	& \sin 2\alpha	& \sin [2(\alpha + \theta_d)]  \\
	\bs
	\optbar{B_d}\ra\psi K_S	& \sin 2\beta	& \sin [2(\beta  - \theta_d)]  \\
	\bs
	B^\pm\ra DK^\pm\ra (f_{1,2})K^\pm  & \cos 2\gamma	& \cos 2\gamma  \\ 
	\bs
	\optbar{B_s}\ra\psi\phi	& \sin 2\chi & \sin [2(\chi + \theta_s)]   \\
	\bs
	\optbar{B_s}\ra D_{s}^{\pm} K^\mp	& \cos 2\gamma & \cos [2(\gamma -
	    2\theta_s)]  \\
	\bottomline
	\end{array} \]
\end{table}

One way to overcome this insensitivity to PBSM is to add a measurement of
$\tilde{\chi} = \chi + \theta_s$ via the fourth process in Table \ref{t2}. To
an accuracy of a few per cent, the true angles $\alpha$, $\beta$,
$\gamma$, and $\chi$ satisy \cite{r8}
\beq  \frac{\sin \alpha  \sin \chi}{\sin \beta  \sin \gamma} = 
		\left| \frac{V_{us}}{V_{ud}} \right|^2  .   \eeq
The right-hand side of this relation is just the square of the Cabibbo angle,
and is very accurately known. Now, for measured angles $\tilde{\alpha}$,
$\tilde{\beta}$, $\tilde{\gamma}$, and $\tilde{\chi}$ which are not the true
angles $\alpha$, $\beta$, $\gamma$, and $\chi$, the relation (\ref{eq4.5}) will
in general fail, even if $\tilde{\alpha} + \tilde{\beta} = \alpha + \beta$ and
$\tilde{\alpha} + \tilde{\beta} + \tilde{\gamma} = \pi$. Thus, a nonvanishing 
$\theta_d$ and/or $\theta_s$ from PBSM would be revealed. \cite{r25}

Another way to try to uncover evidence of PBSM in $B-\overline{B}$ mixing is
to measure $\gamma$, not in the decay $B^\pm \ra DK^\pm$, which does not
involve $B-\overline{B}$ mixing, but in $\optbar{B_s} \ra D_s^\pm K^\mp$,
which does. As indicated in Table \ref{t2}, the latter decay would yield for
$\gamma$ a measured value $\tilde{\gamma} = \gamma - 2\theta_s$. Combining this
$\tilde{\gamma}$ with the $\tilde{\alpha}$ from $\optbar{B_d} \ra \pipm$ and
the $\tilde{\beta}$ from $\optbar{B_d} \ra \psi K_S$ would give
\beq  \tilde{\alpha} + \tilde{\beta} + \tilde{\gamma} = \pi - 2\theta_s  . \eeq
A nonvanishing $\theta_s$, if present, would thereby be revealed.

If $\gamma$ is measured both in $B^\pm \ra DK^\pm$ and in $\optbar{B_s} \ra
D_s^\pm K^\mp$, and a nonvanishing $\theta_s$ is present, different values will
be obtained from the two measurements, uncovering the $\theta_s$. This
illustrates the virtue of making ``redundant'' measurements.\cite{r26,r27}  
  
\section{Summary \label{sect5}}
\setcounter{eqns}{0}
The most obvious candidate for the source of CP violation is the SM weak
interaction. If this interaction is indeed the source, then CP violation comes
from complex phase factors in the CKM matrix. The hypothesis that these phase
factors are the origin of CP violation will be stringently tested in future
experiments, mostly in the $B$ system. Physics beyond the SM could be revealed
by failures of this test. To seek such new physics in CP violation and related
phenomena, we should overconstrain the system as much as possible, being
careful not to restrict the measurements we make to those which could hide the
new physics.



\begin{thebibliography}{99}
\bibitem{r23}The angle $\alpha$ is defined by \Eq{3.1}, $\beta$ by \Eq{3.2} and
$\gamma$ by $\gamma \equiv \arg\, (-V_{ud}V_{ub}^* / V_{cd}V_{cb}^*)$, or
equivalently by $\gamma \equiv \pi-\alpha-\beta$. Thus, the true $\alpha$,
$\beta$, and $\gamma$ always satisfy the constraint (\ref{eq4.3}), even if
unitarity is violated and the $db$ unitarity triangle does not close.
\bibitem{r26}If the tiny branching ratio for $K_L \ra \pi^0\nu\overline{\nu}$
can be measured, we would gain additional clean CKM information which would
help us significantly to test the SM of CP violation. See Buchalla G \etc
Ref.~\cite{r7}
\end{thebibliography}
\end{document}